\documentclass{article}
\usepackage{amsmath}
\usepackage{amssymb}
\usepackage{amsthm}
\usepackage{bbm}
\usepackage[utf8]{inputenc} 
\usepackage{ismir,amsmath,cite,url}
\usepackage{graphicx}
\usepackage{color}
\usepackage{caption}
\usepackage{subcaption}

\title{Assessing the Impact of Sampling, Remixes, and Covers on Original Song Popularity}

\multauthor
{Guilherme Soares S. dos Santos \hspace{1.2cm} Flavio Figueiredo} {
 Universidade Federal de Minas Gerais\\
{\tt\small \{guilhermesoares, flaviovdf\} @dcc.ufmg.br} \\ Dataset at: \url{https://github.com/uai-ufmg/whosampleddata}}

\def\authorname{G. dos Santos and F. Figueiredo}

\begin{document}

\maketitle

\begin{abstract}
Music digitalization has introduced new forms of composition known as ``musical borrowings'', where composers use elements of existing songs—such as melodies, lyrics, or beats—to create new songs. Using Who Sampled data and Google Trends, we examine how the popularity of a borrowing song affects the original. Employing Regression Discontinuity Design (RDD) for short-term effects and Granger Causality for long-term impacts, we find evidence of causal popularity boosts in some cases. \textit{Borrowee} songs can revive interest in older tracks, underscoring economic dynamics that may support fairer compensation in the music industry.
\end{abstract}

\section{Introduction}\label{sec:introduction}

Digitization has drastically transformed the production, distribution, and consumption of music. Advances in data storage, transmission, and processing technologies have brought about significant changes in the music industry, making digital music predominant. This shift has led to new forms of composition, known as ``musical borrowings'', where composers and producers draw on existing musical influences to create new works, thereby expanding the boundaries of musical genres.

An example of musical borrowing is Beyoncé’s 2003 hit ``Crazy in Love'' one of her most iconic songs. This work calls this song a \textit{borrowee}. The \textit{borrowee’s} introduction, marked by triumphant brass, drew significant attention. The brass section was sampled from The Chi-Lites' 1970s song ``Are You My Woman'', here called a \textit{borrowed} song. Created in a pre-digital era, the \textit{borrowed’s} influence on ``Crazy in Love'' is evident, although its precise contribution to the song's success remains difficult to quantify.

We explore three primary forms of musical borrowing: sampling, remixes, and covers. Sampling, a technique popularized in the late 1970s, involves using fragments of pre-existing songs. Widely used in hip-hop, electronic, and experimental music, sampling allows for creative expression through the manipulation and combination of different tracks. On the other hand, remixes involve transforming an existing song by altering its structure, rhythm, or effects, often revitalizing it for new audiences. Finally, covers involve re-recordings that usually offer a fresh interpretation and sometimes even surpass the original in popularity.

Musical borrowings raise significant legal issues, particularly regarding the copyright laws that govern these practices -- debates around fair use often center on whether the borrowing negatively impacts the market for the original work. A study titled ``Sampling Increases Music Sales: An Empirical Copyright Study'' highlights that, in the U.S., the Supreme Court considers the market impact on the original song as a key factor in determining fair use \cite{https://doi.org/10.1111/ablj.12137}. The research suggests that digital sampling can boost sales of the original songs, indicating that musical borrowings may, in some cases, qualify as fair use. Understanding the consequences of using samples in original songs can contribute to fairer legal discussions and more informed decisions by musicians and managers.

Our research aligns with the above discussion, but it also presents counter-arguments, as we now detail. Specifically, we focus on how the popularity of a \textit{borrowee} song (cover, sample, or remix) influences the popularity of the \textit{borrowed} song. Data from WhoSampled\footnote{https://www.whosampled.com}, a crowdsourced website cataloging samples, remixes, and covers, and Google Trends\footnote{https://trends.google.com}, which measures relative search interest, provide the basis for this analysis.

\begin{figure}[t!]
    \centering
    \includegraphics[width=0.95\linewidth]{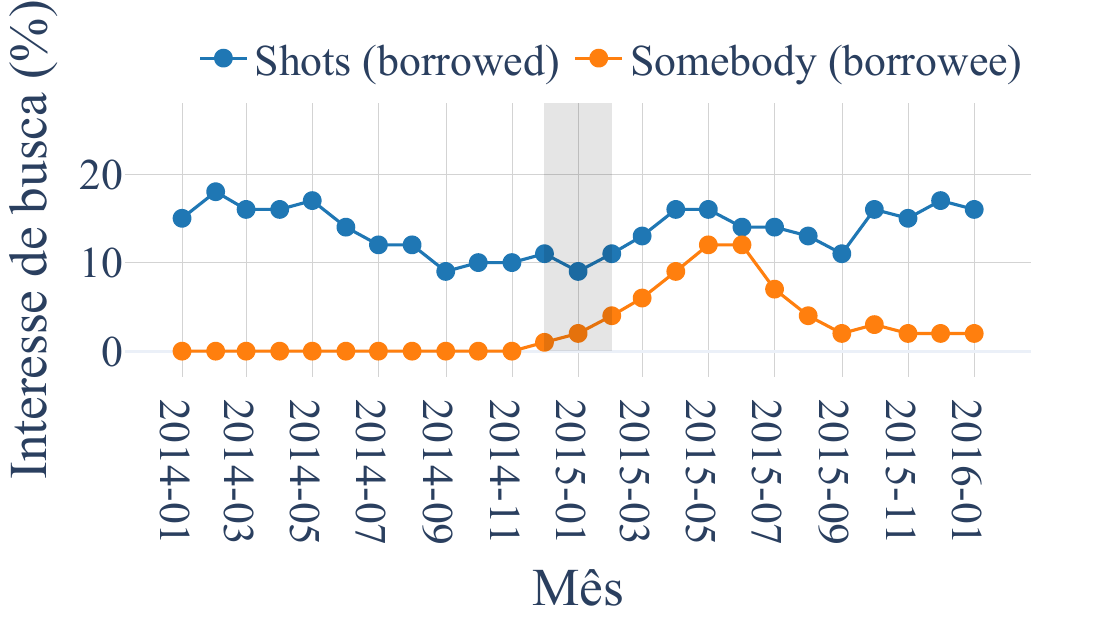}
    \caption{Search Interest Over Time for ``Shots'' (LMFAO) and ``Somebody'' (Natalie La Rose). The release month is the shaded region.}
    \label{fig:ex}
\end{figure}

To effectively measure an effect, we employ two techniques. First, we explore Regression Discontinuity Design~\cite{rdd} (RDD) to measure the immediate causal impact of the \textit{borrowee's} release. Next, we make use of Granger Causality~\cite{granger} as a means to understand the lasting impact of a \textit{borrower} after release. As an example of a \textit{borrowee} that impacts a \textit{borrowed} song, we show in Figure~\ref{fig:ex} the song ``Somebody'' by Natalie La Rose (released in late 2014) that sampled ``Shots'' from LMFAO. Here, we can see that ``Somebody'' renews interest in the LMFAO song.

In general, it is important to note that establishing causality is challenging. Among the approximately 884 instances of musical borrowings analyzed, we identified 182 cases (20\%) showing any causal evidence. From these, our results show that only a fraction of the \textit{borrowees} (45\% or 82 out of 182) have an immediate causal impact via our RDD study. Out of these, 64\% (or 117 out of 182) of these cases show a lasting effect via Granger Causality. It should be emphasized that this analysis focuses on web search interest, rather than sales metrics. Nonetheless, even with a relatively limited dataset, we observed some influence of \textit{borrowees} on the originals. Thus, we hope that our research sheds light on the actual implications of sampling.

In the following sections, we begin with a review of related work (Section~\ref{sec:related_work}), followed by a description of our datasets and methodology (Section~\ref{sec:methodology}) and our results (Section~\ref{sec:effect}). Finally, we conclude with a discussion of our findings and their broader implications (Section~\ref{sec:conclusion}).


\section{Related Work}\label{sec:related_work}

The term ``musical borrowings'' defined by Burkholder \cite{burkholder1994uses}, refers to the techniques composers use to create music based on pre-existing melodies or structures. This practice has been studied not only in music but also in legal contexts, especially regarding copyright. Scholars have highlighted discrepancies between copyright law and the realities of musical composition. Arewa \cite{arewa2005jc} critiques copyright law’s focus on ``romanticized authorship''. This assumes originality and autonomy, disregarding musical borrowing, particularly in genres like hip-hop, where sampling has raised legal concerns.

Musical borrowing also intersects with social inequalities. Hesmondhalgh \cite{hesmondhalgh2006digital} explores how white musicians have historically appropriated works by Black artists, reinforcing cultural disparities. This issue extends to non-hegemonic markets, such as Latin America and Africa, where musical borrowing can have broader implications.

Analyzing musical influence presents challenges due to its subjective nature. Shalit et al. \cite{pmlr-v28-shalit13} tackled this by applying Dynamic Topic Models to a dataset of over 24,000 songs from 1922–2010, revealing that musical influence and innovation are not monotonically correlated. The study identified critical periods of high innovation in the early 1970s and mid-1990s.

Turning our attention to authors exploring a dataset similar to ours, Bryan et al. \cite{bryan2011musical} also explore WhoSampled data to analyze to characterize networks. The study examined centrality in a network of 42,447 samples, applying complex network techniques and power-law distributions to measure the influence of songs and artists. Katz Centrality was introduced as a new method to quantify musical influence, though the author's study did not directly address how sampling affects popularity.

Ortega \cite{ortega2021cover} analyzed cover versions using a network of over 106,000 artists and 855,000 covers from Second-Hand Songs, focusing on the impact of covers on popular music history. Key findings showed genre and language as primary factors in collaboration and influence.

\section{Methodology and Data Collection}\label{sec:methodology}

Our work focuses on analyzing the popularity of songs involved in musical borrowings over time to measure the impact of the release of a new version on the original version. To do so, we used a dataset obtained from Who Sampled containing about 700,000 songs, including samples, covers, and remixes. This dataset was crawled in 2019, and we asked WhoSampled for permission to crawl those data. The website catalogs musical borrowings for more than 975,000 songs from 300,000 artists, contributed by a community of about 28,000 members and verified by moderators or staff\footnote{https://www.whosampled.com/about}. In this sense, our original dataset comprises most of these songs.

The obtained data is structured as a directed graph, where nodes represent songs and edges represent musical borrowings, each labeled to indicate the type of musical borrowing. Since Who Sampled does not provide popularity data or external identifiers, we supplemented the data with Wikidata\footnote{wikidata.org} and Google Trends information.

Initially, we resorted to Wikidata to gather unique identifiers for each song. The song ``Clube da Esquina vol. 2'' by Milton Nascimento has a Wikidata identifier of \texttt{Q20053386}. More importantly, on Wikidata, this song also has its Freebase\footnote{A discontinued ontology that is still used internally by Google Trends -- https://en.wikipedia.org/wiki/Freebase\_(database)} identifier listed: \texttt{/m/0zjw3z\_}. This Freebase identifier was later used for Google Trends queries. Before describing how we collected time series, we discuss how we gathered the Freebase id of our songs.




To find song identifiers, we used Wikidata with three search queries: one with the song title, one with the artist name, and one combining both. For each, we retrieved the top ten entities, removed duplicates, and processed entities in batches of 50 to extract relevant properties.

Several steps were taken to gather the correct IDs, as now detailed. Initially, it is essential to consider that some entities returned may not strictly represent a musical work during the textual search using the song and artist name. To address this issue, we filtered entities using the \textbf{``instance of (P31)''} property in Wikidata, selecting those classified as \textbf{``musical work (Q2188189)''} or subclasses like \textbf{``song (Q7366)''} or \textbf{composition (Q204370)''}. This approach allows us to filter out entities not directly related to musical works.

From the original 699,123 songs, this filtering reduced the list to 111,238 entities, of which 66,373 (59.6\%) lacked knowledge base identifiers. Focusing on entries with Freebase IDs, we finalized a set of 44,857 songs.

To ensure collected entities matched songs in our dataset, we refined them based on the normalized string similarity between the song and artist names, calculated separately with the \texttt{difflib} library\footnote{https://docs.python.org/3/library/difflib.html} in Python. We retained only entries with a string distance above 0.55 for both names, reducing our dataset from 579,322 to 25,830 songs, representing 4,477 musical borrowings.

With the Freebase identifiers, we accessed Google Trends using each identifier as a URL parameter\footnote{https://trends.google.com/trends/explore?q=/m/0zjw3z\_}. Our queries were configured to retrieve the global search interest over time for each song, starting from the inception of Google Trends in 2004. We successfully retrieved data for 4,360 songs, each represented as a monthly time series indicating relative popularity, normalized to peak at 100.

Estimating the causal effects of temporal data is challenging~\cite{hausman2018regression}. Initially, we can point out that several events may impact popularity (e.g., a song may be sampled twice). Furthermore, other seasonal effects (e.g., a good year for an artist) may be in place. To reduce this impact, all of our time series were filtered to have 24 points only, one per month. Twelve of these occurred before the release and Twelve after. Thus, we shall look at the impact on monthly interest. Moreover, we observed that some time series contained zero values in all time intervals, indicating the absence of search interest data for these musical entities throughout the analyzed period. After this last filter, we were left with 884 borrowings.


After this pre-processing step, we calculated the impact of the launch on the original in the short and medium term using RDD and Granger Causality.

\section{Effect of Musical Borrowings on Interest in a Song}\label{sec:effect}

To assess the impact of samples and re-recordings on audience interest in original music, we explored the hypothesis that these elements introduce discontinuity in nature. Considering that the original music already existed before and so did audience interest, we compared interest in the original music before and after the re-recording. This allows us to estimate the impact of the borrowing on audience interest in the original music.

\subsection{Short-term impact}

We used the Regression Discontinuity Design (RDD) method to measure the impact of a sample on the original music. RDD is a statistical method that allows us to estimate the causal effect of a treatment on an outcome variable when the probability of treatment changes discontinuously around a cutoff point. In this case, the treatment is the release date of the \textit{borrowee} song. The outcome variable is the search interest in the original song, the \textit{borrowed}. The cutoff point separates the observations into control and treatment groups. In detail, our RDD was as implemented as follows:
\begin{equation}
googletrends(t) \sim 1 + t + \mathbb{I}_{t>0} + t\cdot \mathbb{I}_{t>0}
\end{equation}

In this context, $t$ represents our time variable, shifted so that the release time of the \textit{borrowee} occurs at time zero. This transformation is commonly used in RDD as it simplifies the interpretation of model parameters. The term $\mathbb{I}_{t>0}$ is an indicator function that equals one when $t > 0$ (after the release) and zero otherwise. Therefore, at $t=0$, this term's weight reflects the difference in intercepts, known as the Average Treatment Effect (ATE), which serves as a causal estimate of the release's impact. We express this measurement in relative terms, where 100\% indicates a doubling of the intercept.

\begin{figure}[t!]
    \centering
    \includegraphics[width=.92\linewidth]{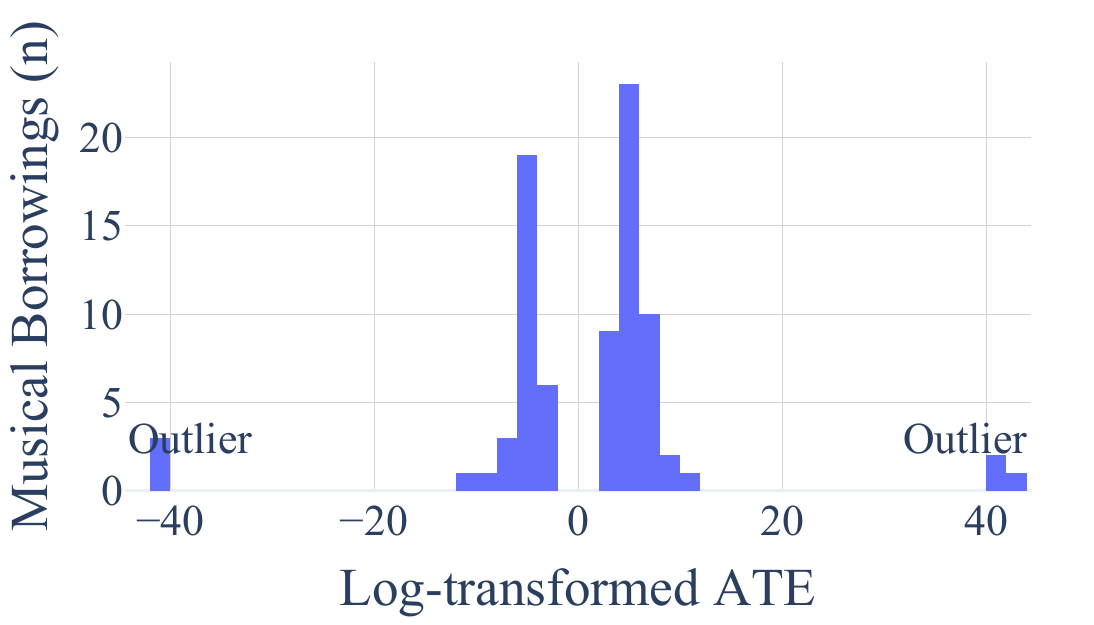}
    \caption{Distribution of the log-transformed average treatment effect (ATE) for the 82 statistically significant musical borrowings.}\vspace{-1em}
    \label{fig:ate_distrib_log}
\end{figure}

Most importantly, we analyzed the ATE estimates only for regressions that achieved statistical significance at a level below 0.05. Our final dataset included 884 musical borrowings, of which 82 (approximately 9.3\%) yielded statistically significant \(p\)-values. Figure \ref{fig:ate_distrib_log} illustrates the ATE distribution for these 82 significant borrowings, which we will discuss next.



\begin{figure*}
    \centering
    \begin{subfigure}[b]{0.32\textwidth}
        \includegraphics[width=\linewidth]{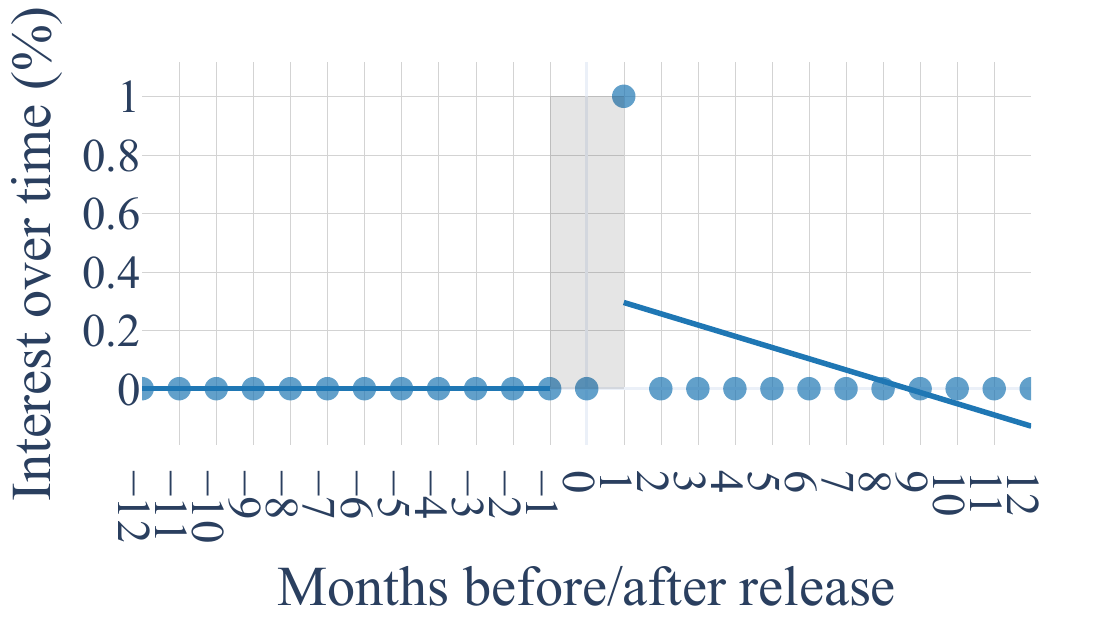}
        \caption{RDD predictions and observed values for the song ``Something's Got a Hold on Me'' by Etta James (outlier).}
         \label{fig:outlier}
    \end{subfigure}
    \begin{subfigure}[b]{0.32\textwidth}
        \includegraphics[width=\linewidth]{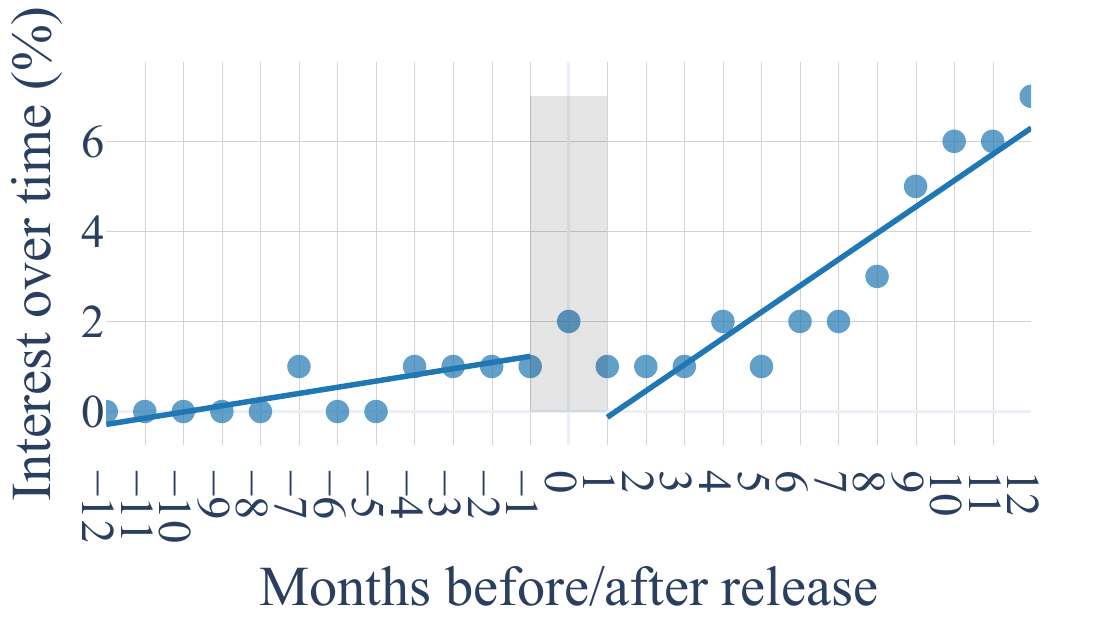}
        \caption{``Summertime Sadness'' (Lana Del Rey, 2012) borrowed by ``Body Electric'' (Lana Del Rey, 2012).}
        \label{fig:66}
    \end{subfigure}
    \begin{subfigure}[b]{0.32\textwidth}
        \includegraphics[width=\linewidth]{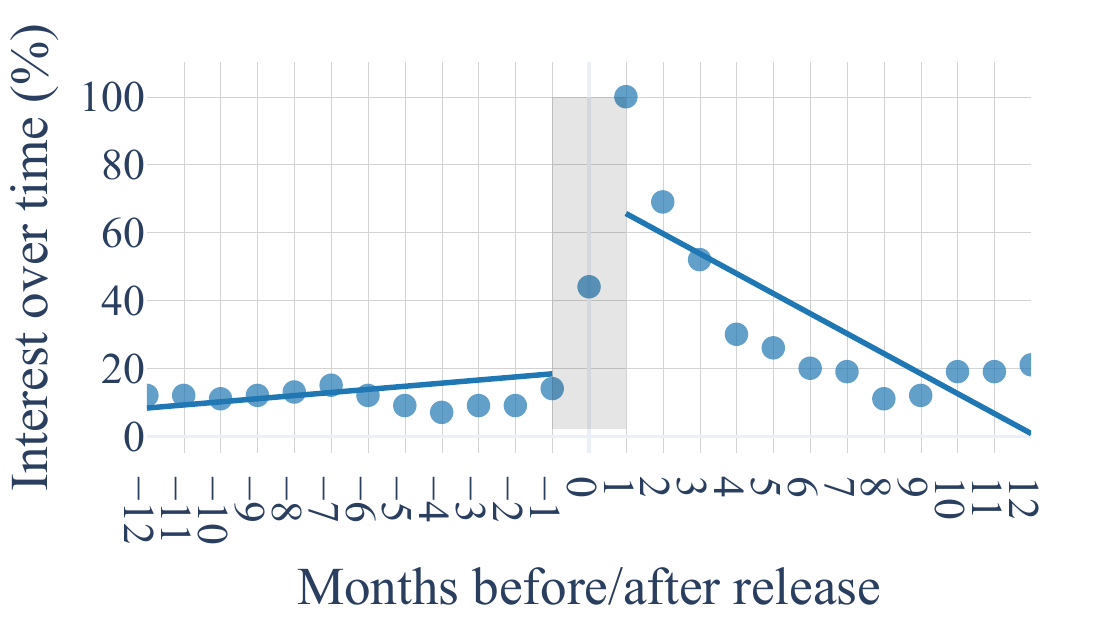}
        \caption{``You Spin Me Round (Like a Record)'' (Dead or Alive, 1984) borrowed by ``Right Round'' (Flo'Rida, 2009)}
        \label{fig:65}
    \end{subfigure}
    \begin{subfigure}[b]{0.32\textwidth}
        \includegraphics[width=\linewidth]{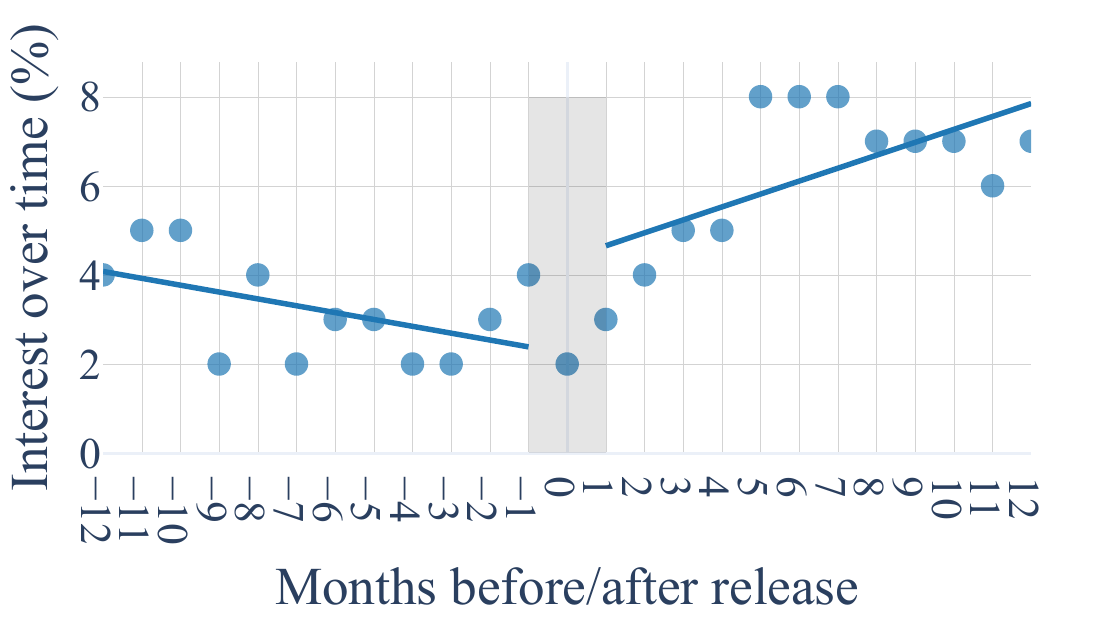}
        \caption{``It Takes Two'' (Rob Base DJ E-Z Rock, 1988) borrowed by
``Peaches N Cream (Snoop Dogg song)'' (Snoop Dogg, 2015)}
        \label{fig:1}
    \end{subfigure}
    \begin{subfigure}[b]{0.32\textwidth}
        \includegraphics[width=\linewidth]{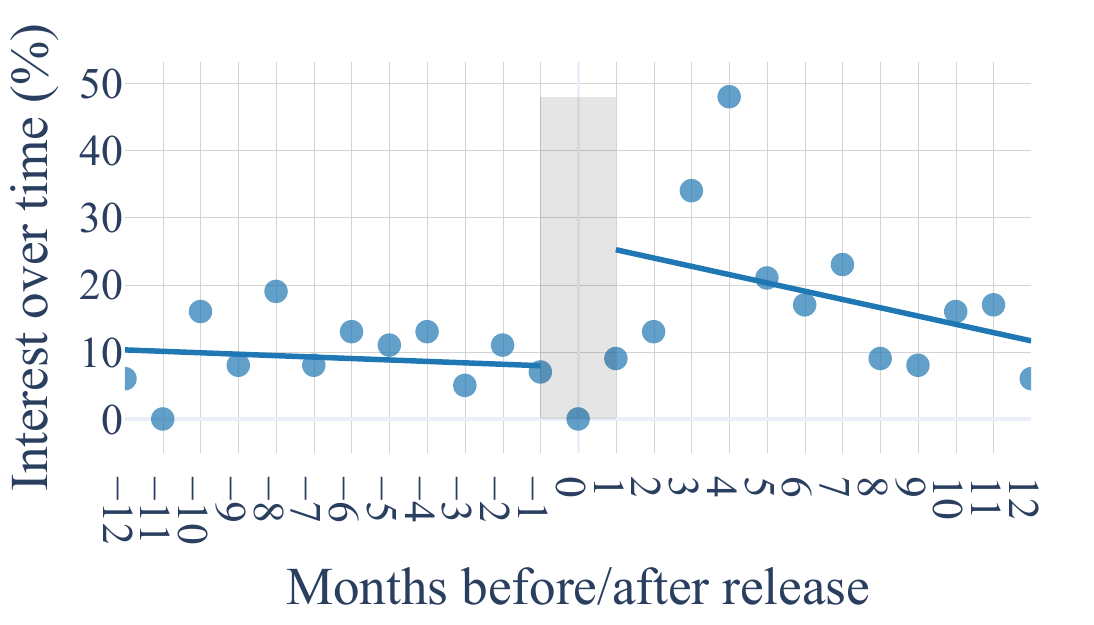}
        \caption{``Close to Me'' (The Cure, 1985) borrowed by the song ``So Human'' (Lady Sovereign, 2009)}
        \label{fig:21}
    \end{subfigure}
    \begin{subfigure}[b]{0.32\textwidth}
        \includegraphics[width=\linewidth]{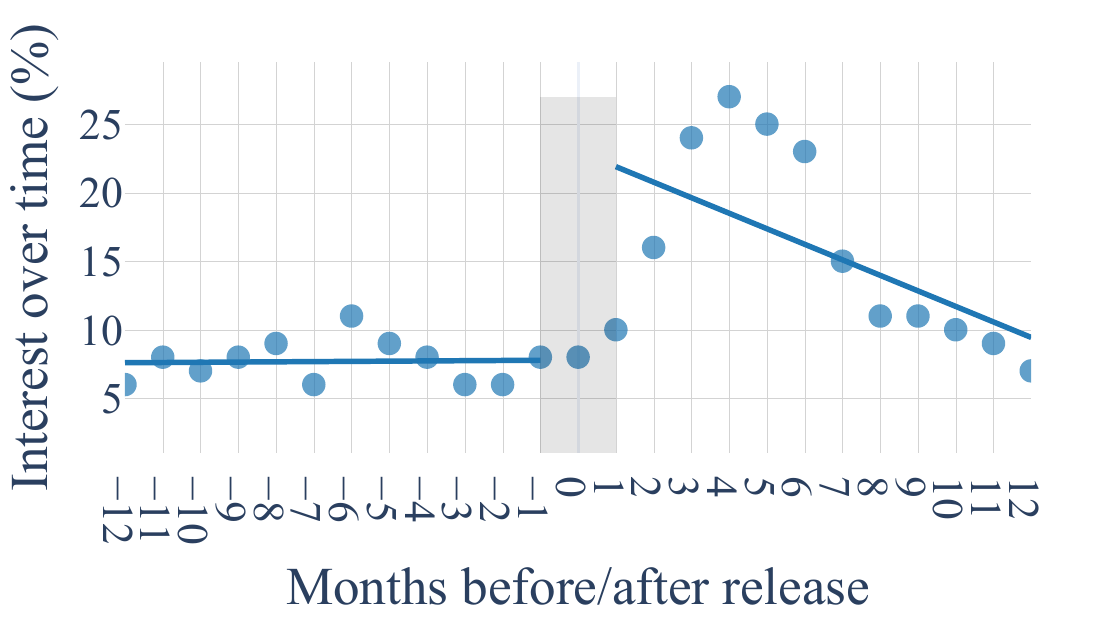}
        \caption{``Werewolves of London'' (Warren Zevon, 1978) borrowed by All Summer Long (Kid Rock, 2007)}
        \label{fig:15}
    \end{subfigure}
    \caption{ATE Examples}
\end{figure*}

We initially point out borrowings with very high or very low ATEs from the figure. In some cases, this occurred because the original song had no prior search activity. However, after the sample's release, the original song began to receive increased search interest. Figure \ref{fig:outlier} shows an example of this phenomenon. The song with the highest ATE in our set is ``Something's Got a Hold on Me'' by Etta James, released in 1962, sampled in 2011 by American rapper Flo-Rida in his song ``Good Feeling''. We manually inspected these extreme events; all were outliers like this one and we disregarded them. 

We thus decided to focus on the non outlier cases, shown in the center of Figure~\ref{fig:ate_distrib_log}. We begin with an interesting case of a negative ATE. Thus, Figure \ref{fig:66} presents the RDD analysis of Lana Del Rey's ``Summertime Sadness,'' sampled in ``Body Electric,'' released five months later on the EP Paraíso. The ATE of -152\% indicates a negative impact on interest in the original song following the sample release. We believe that the EP's marketing and the similarity between the two tracks may have influenced the original song's performance during that period.

Figures~\ref{fig:65},~\ref{fig:1},~\ref{fig:21} and ~\ref{fig:15} all show examples of positive ATE. It is quite interesting that several of these examples show samples reviving interest in songs released in the late '70s to late '80s. This shows evidence that, in some cases, a \textit{borrowee} may revive interest even in older songs. Nevertheless, we still need to check if this revival is long-lasting. Figures~\ref{fig:65} and~\ref{fig:15} show a sharp decay a few months after the \textit{borrowee's} release.  This is why we complement this analysis with our Granger Causality analysis next.

\subsection{Long-term impact}


Granger causality is a statistical technique that investigates whether the past of one time series contains valuable information to predictthe future of another. Thus, it considers the lags between observations.

We use the \textit{statsmodels}\cite{seabold2010statsmodels} library to perform the Granger causality test, which offers robust statistical tools for time series analysis. The test involves formulating autoregressive (AR) models and comparing them to see if including lagged terms from one time series improves the prediction of another time series.

When applying the Granger test, it is necessary to specify the \textit{max\_lag} parameter, determining the maximum number of lags considered. We set the value of \textit{max\_lag} to 10, assuming that the causal relationship between the original song and the musical borrowing can be captured within 10 months (from one month to almost a year of maximum lag). 
The test is performed by combining two time series into a complete data frame and running the statistical procedure. The result of the Granger test is a $p$-value associated with the F-test of the sum of squares of the residuals (SSR F-test) for each lag up to the specified \textit{max\_lag}. 

A low $p$-value (usually less than 0.05) indicates that the musical borrowing time series contains significant predictive information about the original song time series. When this occurs, we have evidence of a Granger Causal relationship for the time series pair.

After the analysis, we observe that 64\% (117) of the musical borrowings have a $p$-value below 0.05, establishing that, in most cases, the time series of the new version (cover, sample, or remix) contains significant predictive information about the time series of the original song. This suggests that \textit{borrowees} generally influence \textit{borrowers} over more extended periods of time.



\section{Conclusions}
\label{sec:conclusion}

This paper estimated causal evidence that musical borrowings impact the popularity of \textit{borrowed} songs. Our motivation was delving deeper into the argument that borrowing (samples, remixes, and covers) increases the popularity of \textit{borrowed} songs.

Our findings indicate some causal relationships between musical borrowings and the popularity of \textit{borrowed} songs, though effects vary in strength. When present, a \textit{borrowee} song can revitalize interest in older tracks, as shown by RDD analysis. Notably, sustained impacts assessed via Granger Causality suggest borrowings can have a lasting influence on original song popularity.


Our work has performed extensive care in collecting an accurate dataset that links several databases (Freebase, MusicBrainz, and Google Trends). Nevertheless, this does not come with some drawbacks. Out of approximately 700,000 songs analyzed, only 884 instances of musical borrowing were identified. Future work could refine data.

Finally, our results shed light on the complex economics of borrowings, with implications for fairer compensation practices~\cite{claflin2020get}.

\bibliography{ISMIRtemplate}

\end{document}